\documentstyle[prl,aps,preprint,psfig]{revtex}

\begin{document}

\title{Nanomechanical resonators operating as charge detectors in the
nonlinear regime}
\draft
\author{H. Kr\"ommer, A. Erbe, A. Tilke, S. Manus, and R.H. Blick$^*$}


\address{
         Center for NanoScience and Sektion Physik,
Ludwig-Maximilians-Universit\"at, Geschwister-Scholl-Platz 1,
80539 M\"unchen, Germany. \\
        }
\date{\today}
\maketitle

\begin{abstract}
We present measurements on nanomechanical resonators machined from
Silicon-on-Insulator substrates. The resonators are designed as freely
suspended
Au/Si beams of lengths on the order of 1~-- 4~$\mu$m and a thickness of
200~nm. The beams are driven into nonlinear response by an applied
modulation at radio frequencies
and a magnetic field in plane. The strong hysteresis of
the magnetomotive response allows sensitive charge
detection by varying the electrostatic potential of a gate electrode.
\end{abstract}

\vspace*{0.8cm}

The mechanical vibration of a violin string produces
audible sounds in the frequency range of some 100~Hz to several 10~kHz.
Halving the length of such a clamped string the eigenfrequencies are
increased by an octave.
Scaling down the string to only some 100~nm yields frequencies in the radio
frequency (RF)
range. Recent work on such nanomechanical
resonators~\cite{cleland2,carr98:3821}
demonstrated different schemes of detection
and their versatility for applications in
metrology~\cite{cleland1,erbe,weiss}.
Integrating mechanically flexible structures with single electron devices
on the nanometer scale offers not only high speed of operation but also broad
tunability of the tunnel contacts. Applications of mechanical
resonators in nonlinear oscillators~\cite{greywall,yurke} or parametric
amplifiers~\cite{rugar} are of great importance for scanning probe
measurements and accurate frequency counters or clocks in general.

In this work, we want to demonstrate how
to build nanometer sized mechanical resonators and how to apply the nonlinear
response of these devices
for charge detection at several 10~MHz. The resonators are operated
in the RF regime with typical dimensions of only a few 100~nm in width
and height. Applying a sufficiently large excitation amplitude, the suspended
beam shows a highly nonlinear response, which in turn allows extremely
sensitive charge detection. Moreover, the device represents a model
to studying phenomena such as stochastic resonance and
deterministic chaos in a mechanical system on the nanometer scale.

The material employed is a commercially available Silicon-on-insulator
(SOI) substrate
with thicknesses of the Si-layer and the sacrificial layer of 205~nm and
400~nm,
respectively.
Processing of the devices requires optical lithography in a first step by
which a metallic Al/Au~(180nm) mask is deposited.
Adjacently the nanostructure is defined by electronbeam lithography
and deposition of an Al/Au-layer with a total thickness of typically
80~nm.
The metal layers deposited on Si during lithography are a thin adhesion layer
of Ni/Cr (1.5~nm), a covering Au-layer (50~nm), and an Al-etch
mask (30~nm). The sample then is dry-etched in a reactive-ion etcher (RIE)
in order to
obtain a mesa structure with clear cut walls. Finally, we perform a
hydro-fluoric
wet-etch step to remove the sacrificial layer below the resonators.

The suspended resonator is shown
in Fig~1: This particular beam has a length of $l = 3$~$\mu$m, a width of $w
= 200$~nm and
a height of $h = 250$~nm and is clamped on both sides. The gate contact
couples on the complete length of the resonator.
The remaining beam is a Au/Si hybrid for
which different elastic moduli have to be taken into account in the numerical
simulations (not shown here). From these and the measurements shown below we
can conclude that the HF attacked the Si-beam slightly, hence the
resonance frequency is lower than usual. This enhances the nonlinear
response of the
beam, since the restoring forces are less rigid. This assumption is
verified by the close-up
of the suspended beam shown in the inset of Fig.~1.

All measurements were conducted at 4.2~K in a sample holder with a residual
$^4$He-gas pressure of about $10^{-2}$~bar. This ensures thermal coupling,
while it naturally reduces
the mechanical quality factor (defined as $\kappa = f/ \Delta f$, where $f$
is the
frequency at resonance and $\Delta f$ half width of the maximum).
The sample is mounted between interconnecting microstrip lines,
designed to feed the circuit with
frequencies up to 10~GHz, and a magnetic field is applied
perpendicular to the beam. The absolute resistance of the metal wire on top of
the
resonator was found to be 34~$\Omega$, which results in an appropriate
impedance matching of the circuit.
The beam is set into motion by applying a high frequency electromagnetic
excitation in the range of 10 -- 100~MHz and ramping a magnetic field in
plane. This
results in an effective Lorentz force generated perpendicular to the sample
surface.
For excitation and final amplification we use an HP~8594A spectrum analyzer.
The hysteresis of the mechanical resonator is probed with an additional
Marconi~2032
source, which can be ramped from lower to higher frequencies and vice versa.
The preamplifier employed is a low-noise broad-band (UHF- to L-band)
JS amplifier (MITEQ Corp. 1998)
with a specified noise figure of $NF = 0.6$~dB and gain of $G = 30$~dB.

The capacitive coupling between beam and gate (see inset Fig.~1)
is determined by numerical
evaluation with a commercially available program
(MAFIA, electromagnetic problem solver, ver.~3.20).
>From these calculations we obtain a capacitive coupling between gate and beam
in the linear regime of
$C_{gb} \cong 220$~aF. The frequency shift $\delta f$ of the mechanical
resonance results from the capacitive coupling given by
the electrostatic energy $E = Q^2/2C$, where $Q$ is the accumulated
charge and $C$ the capacitance between resonator and gate.
This term can be expanded with regard to the elongation amplitude $u =
u(t)$ of
the suspended beam, which yields for the electrostatic energy
with $C = C(u) $ via a truncated Taylor expansion

\begin{equation}
E (u) = \frac{1}{2} \frac{Q^2}{C} \cong \frac{1}{2}
                     \frac{Q^2}{C + \frac{1}{2} C'' u^2}
     \cong \frac{1}{2} \frac{Q^2}{C} \left( 1 - \frac{1}{2} \frac{C''}{C} u^2
                                     \right)
     = E - \frac{1}{4} \frac{Q^2}{C^2} C'' u^2,
\end{equation}

where $C'' = \frac{\partial^2 C(u)}{ \partial u^2}|_{u=0}$ represents the
second derivative of the
capacitance with respect to the spatial coordinate $u$ at $ u = 0$.
This gives with $Q = CU$ a frequency shift of the mechanical resonance on
the order of

\begin{equation}
\delta f = \sqrt{f^2 -  \frac{C''}{2m_{eff}} V^2} - f
              \cong - \frac{C''}{4m_{eff} f^2} V^2,
\end{equation}

where $m_{eff}$ is the beam's effective mass (in our case $\sim 4.3 \times
10^{-16}$~kg) and $V$ the applied gate voltage. It has to be noted that
for an absolute charge measurement the necessary charging of all metallic
contacts, e.g. bond pads and leads have to be taken into account. For one
of the bond pads for example we estimate a capacitance of
$C_{bp} = \epsilon A/d \cong 2.11$~fF.
However, it is still possible to determine the relative charge $\delta q$
on the
closely connected gate with a high accuracy as will be shown below.

In Fig.~2(a) the RF response of the beam is depicted:
Applying a magnetic field in plane, we find an increase of the
peak maximum proportional to $\sim B^2$ (plotted in
the inset). The driving amplitude of the RF is $-66$~dBm, ensuring linear
response of the
suspended beam.
The FWHM of the resonance is $\Delta f = (16 \pm 0.2)$~kHz, resulting in
a mechanical
quality factor at resonance of $\kappa = f_0 / \Delta f = 2330 \pm 30$. We
find
a large
discrepancy compared to macroscopic mechanical oscillators with $\kappa \sim
10^5$~\cite{greywall}.
This can be explained by the coupling gas in the sample holder and the fact
that the
surface tension in these small devices naturally has a larger influence
than in
macroscopic systems.

In Fig.~2(b) the power coupled into the resonator is increased from
$-70$~dBm to $-50$~dBm where we find a strong nonlinear response.
In the present case the nonlinear response is identified by the distorted peak
shape. Above a critical value of the excitation voltage the curve finally
shows
a bistability accompanied by a pronounced hysteresis.
The transition occurs at
about $-53$~dBm, although an asymmetry of the peak structure is found at
$-59$~dBm
already. The nonlinearity is caused by the variation of the restoring force at
the clamping points~\cite{cleland2} and can be modelled by adding a cubic term
in the equation of motion of the beam~\cite{greywall}. Comparing   our data
with a model derived earlier \cite{yurke} we find excellent agreement
(modelled
traces are not shown).

A closer inspection of the nonlinear response seen in Fig.~2(b) can be
obtained
with an external frequency source, while monitoring the absorption on the
spectrum
analyzer. This allows to approach the hysteretic region around $f =
37.287$~MHz
and 37.294~MHz from larger and lower frequencies. In Fig.~3 such a
measurement is
shown: The inverted triangle ($\bigtriangledown$) corresponds to an
increasing frequency, while the
triangle ($\bigtriangleup$) represents the lowering branch.
The applied power is $P_{exc} = -49$~dBm
and the magnetic field $B = 12$~T. Within this bistable region
(width $\Delta f_{hys} \sim 7$~kHz) the resonator is very sensitive to charge
fluctuations on the nearby gate.
Following this idea the suspended beam is a prime candidate to study
stochastic resonance~\cite{gammaitoni} in a nanomechanical resonator at RF.

Optimum operating conditions are obtained by fixing the driving amplitude at
the critical point with maximum slope (traces in Fig.~4). The excitation
power is
levelled at $-52.8$~dBm and the magnetic field at 12~T. As seen in the inset
the peak
position varies as the square of the gate voltage applied. We achieve a
sensitivity of
$\Delta V / \sqrt{\Delta f} \cong 4.1 \times 10^{-2}$~V$/ \sqrt{\mbox{Hz}}$.
The slope at the critical point $dA/d f|_{f = f_c} \rightarrow \infty$
diverges, resulting in extremely sensitive amplification.
In the measurements presented in Fig.~4 we obtain a charge resolution at a
finite bias on the gate ($V = \pm 4$~V) of
$\sim 0.7 \times 10^2$~e/$\sqrt{\mbox{Hz}}$ limited by electronic noise.
It is important to note the enhancement of sensitivity with increasing
gate voltage (see inset of Fig.~4).
The accuracy of the measurement can be further enhanced by determining the
phase shift
of the resonator, as shown in Fig.~5. For this measurement we modified our
arrangement according to Ref.~\cite{cleland1}, i.e. including a mixer and a
phase
shifter.
With this setup it was possible to obtain a sensitivity of
$\sim 1.0 \times 10^{-1}$~e/$\sqrt{\mbox{Hz}}$. As before the operating
point is
adjusted in the transistion region at the critical point (as indicated in
Fig.~5). Biasing the resonator to $V = \pm 4$~V, we obtain a charge
resolution of
$\delta q = n e = 70$~$e$.

In summary, we demonstrated a new method to apply nanomechanical resonators as
charge detection devices at radio frequencies. The main features are the
high speed
of operation and the increased sensitivity, due to the operation in the
nonlinear regime.
An interesting outlook is given by considering the quantum limit for the
resonators
applied in this work: Assuming a bath temperature of $T_b = 10$~mK and
computing the relation $k_B T_b/h = f_{mech}$, we obtain a
frequency limit of about 200~MHz for the quantum
mechanical oscillation regime (with $k_B$: Boltzmann constant; $h$: Planck's
constant).

We like to thank J\"org P. Kotthaus for his support and
for helpful discussions. Also we thank Klaus Richter and Richard J.
Warburton for
critically reading the manuscript. This work was funded in part by the
Bundesministerium f\"ur
Forschung und Technologie (BMBF) and the Deutsche Forschungsgemeinschaft
(DFG). \\

$^*$: Corresponding author: Robert H. Blick / Electronic mail:
robert.blick@physik.uni-muenchen.de



\newpage

\figure{Fig.~1: 
Micrograph of the nanomechanical resonator used in the experiment: The
gate couples to the resonator, machined out of Si with a
50~nm evaporated Au-layer. Inset: Close-up of the resonator applied
in the measurements.
       }
\label{one}

\figure{Fig.~2: 
(a) Mechanical resonance at $f = 37.26$~MHz: Excitation level is fixed at
$-66$~dBm with peak maximum increasing as $B^2$ (see inset).
(b) Beam response driven into the nonlinear regime at $B = 12$~T with
power levels increasing as indicated from $-70$ to $-50$~dBm.
       }
\label{two}

\figure{Fig.~3: 
Hysteretic response of the excited beam: The inverted triangle
($\bigtriangledown$)
indicates the signal with rising frequency, while the triangle
($\bigtriangleup$)
represents lowering of the frequency ($P_{exc} = -49$~dBm and $B = 12$~T).
       }
\label{three}

\figure{Fig.~4: 
Operating the resonator in the transition region at $-52.8$~dBm
with maximum signal sensitivity. Resonance traces are shifted by an applied
gate voltage. Note the shifting of the critical point when the gate voltage
is varied (inset).
       }
\label{four}

\figure{Fig.~5: 
Measured phase variation of the resonator in the transition region as in
Fig.~2(b).
Inset: Phase of the resonator at
      a fixed gate voltage.   
       }
\label{five}











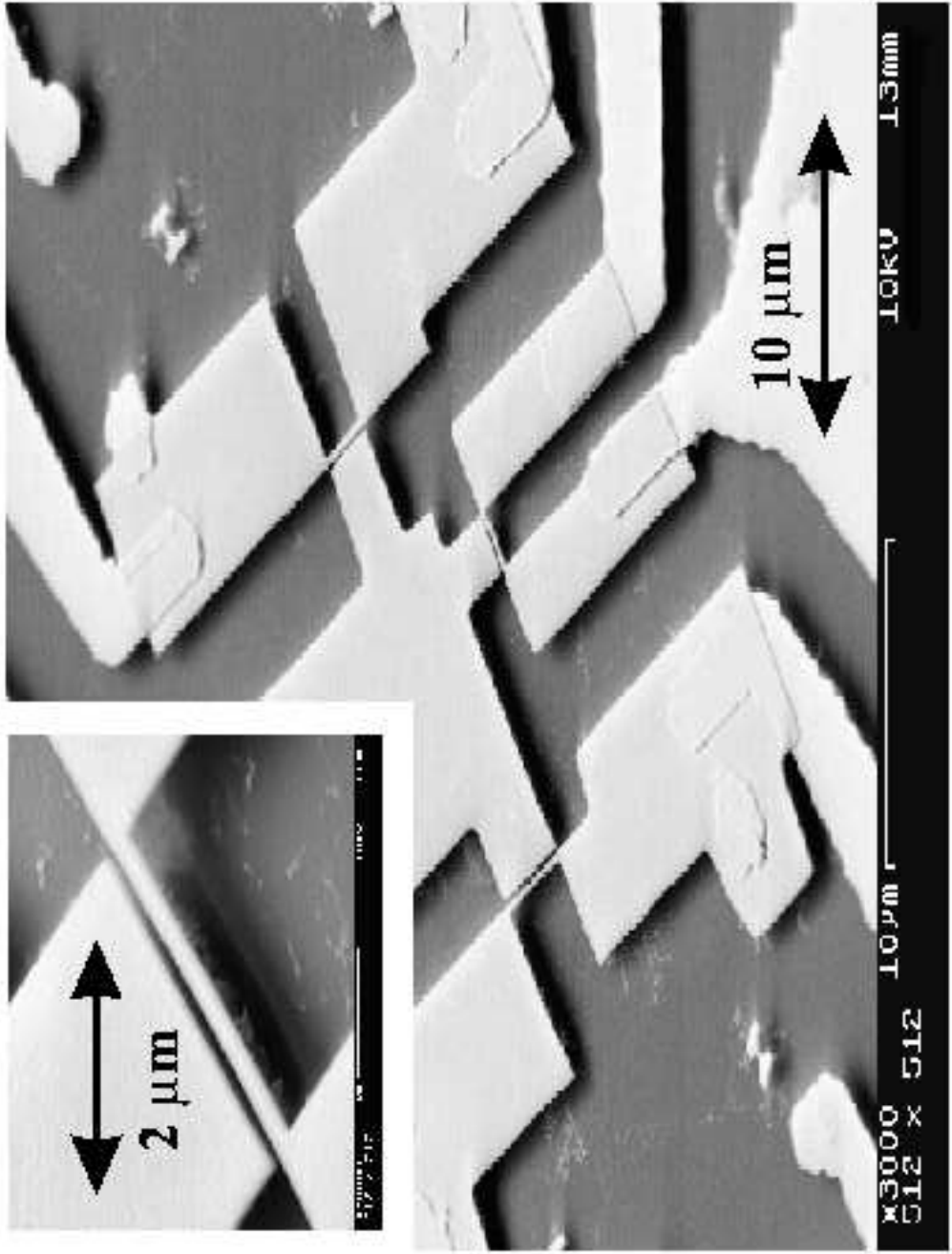
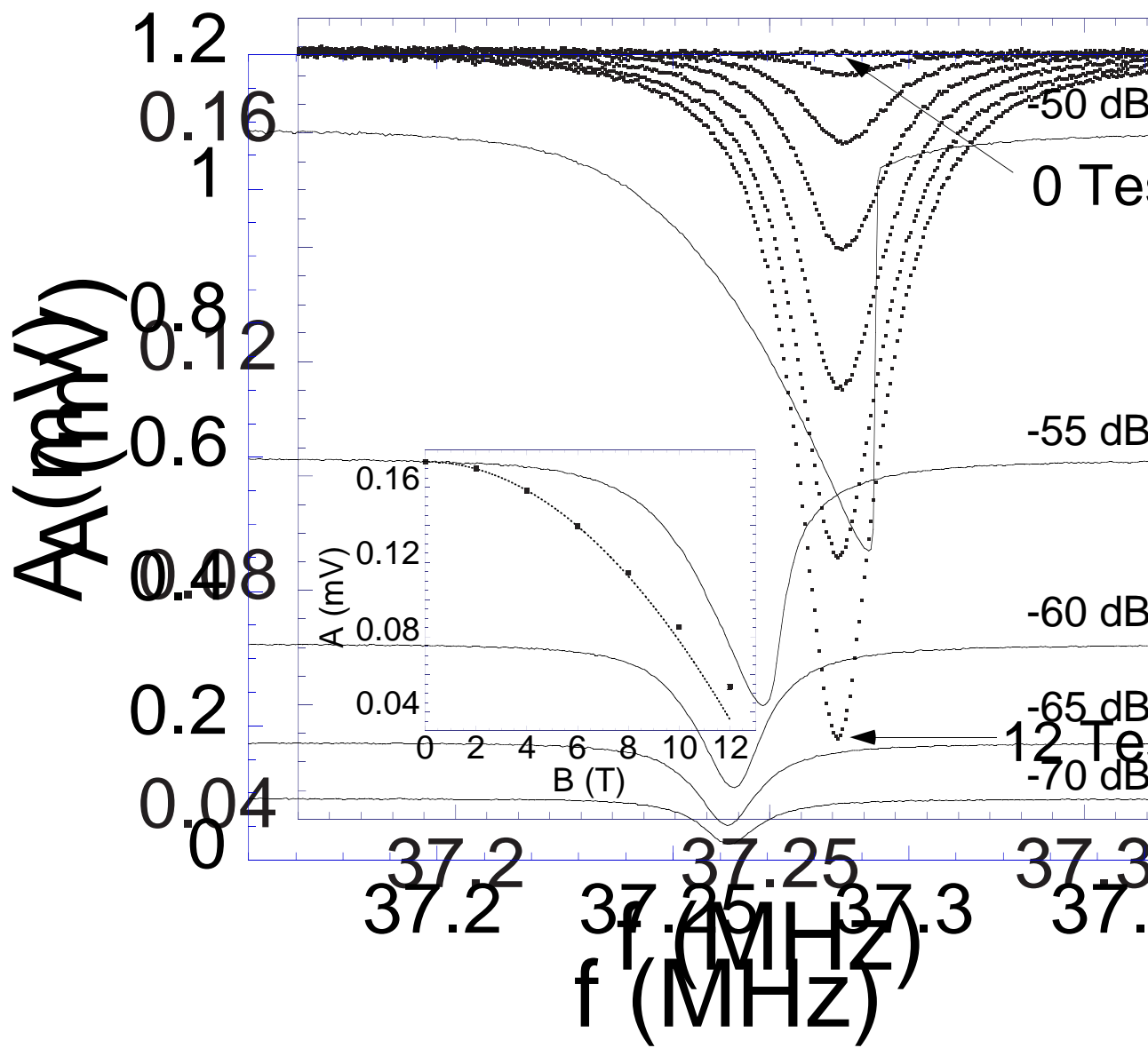
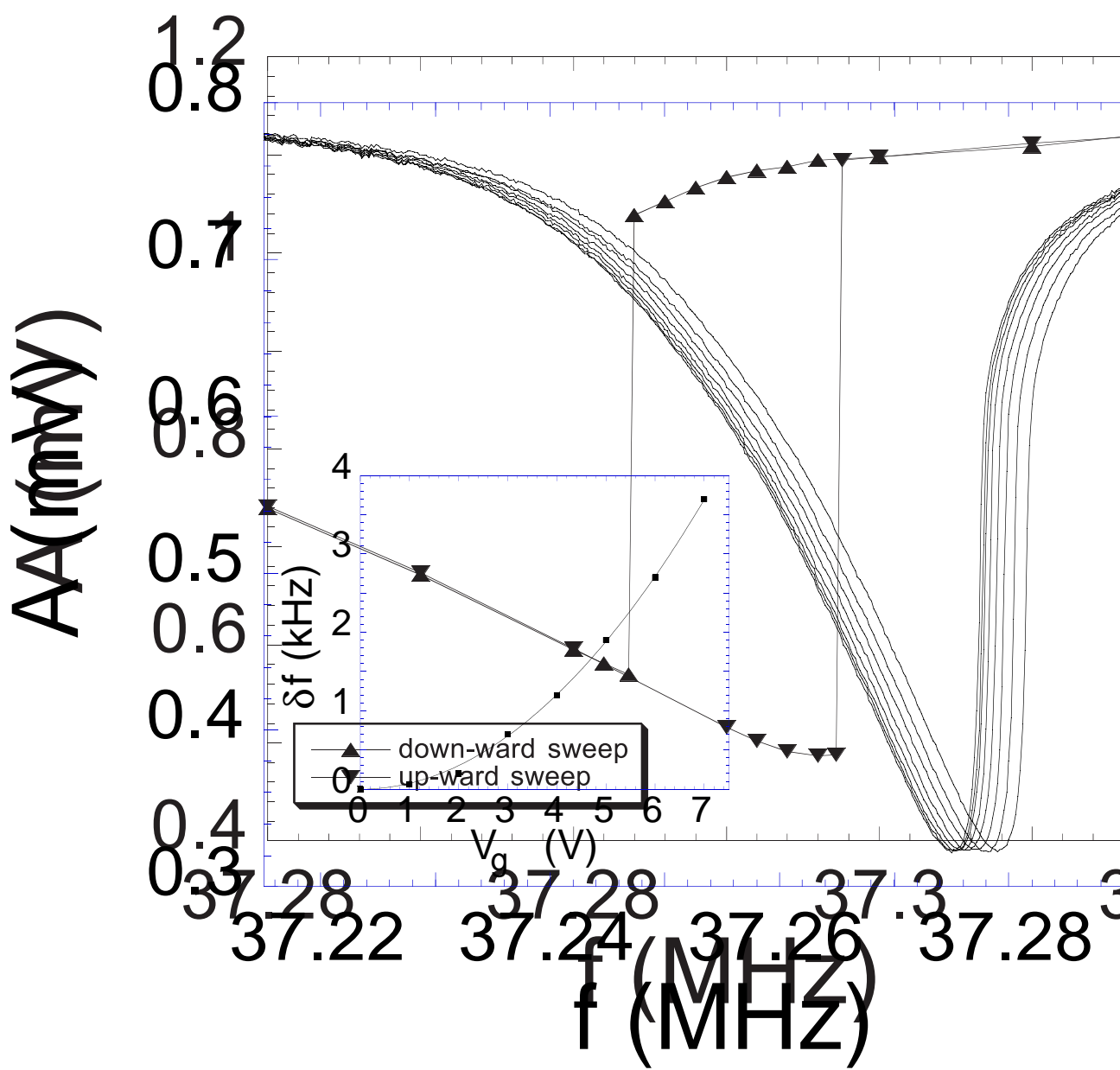
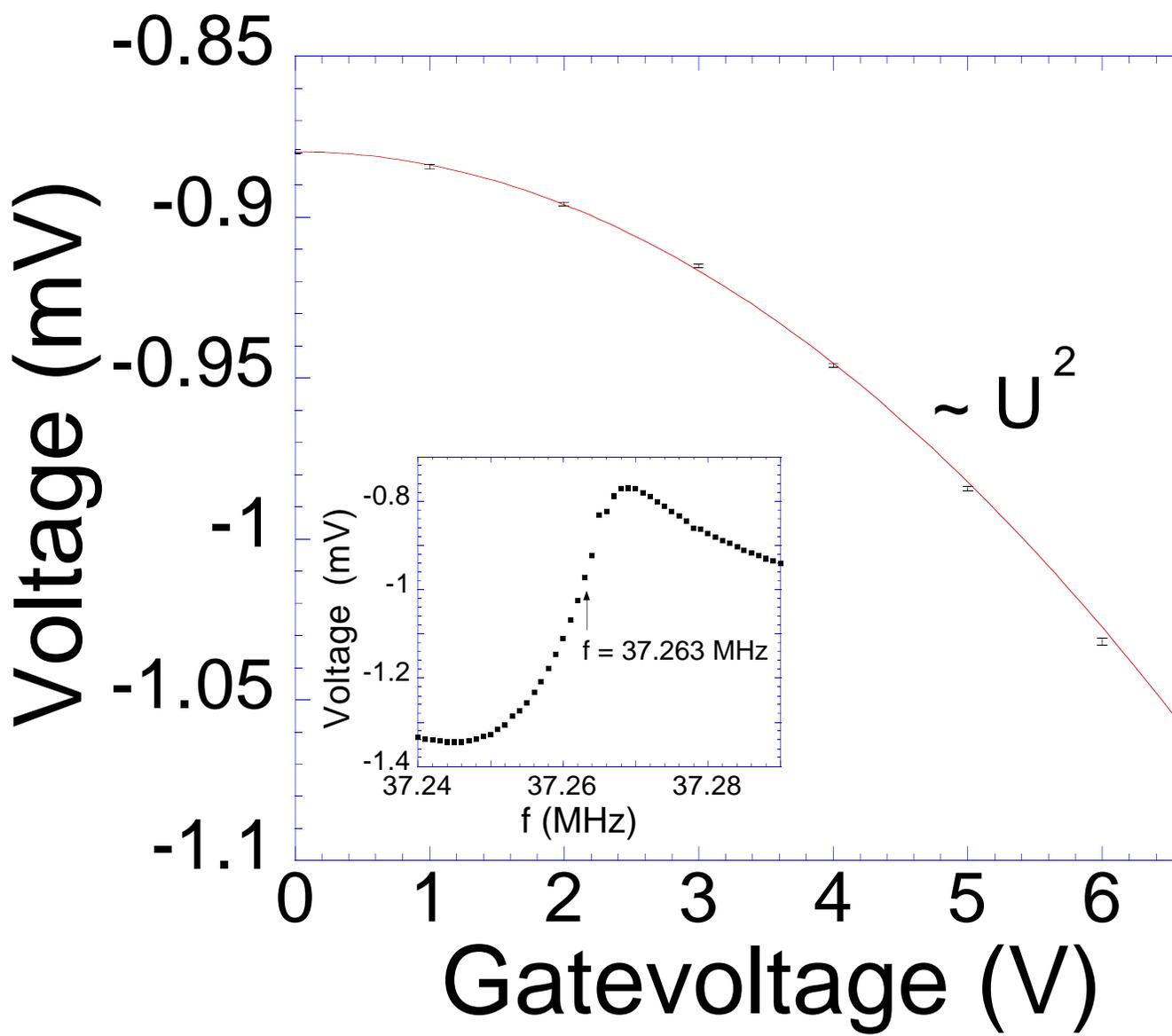
\end{document}